
\newskip\oneline \oneline=1em plus.3em minus.3em
\newskip\halfline \halfline=.5em plus .15em minus.15em
\newbox\sect
\newcount\eq
\newbox\lett

\def\simlt{\mathrel{\lower2.5pt\vbox{\lineskip=0pt\baselineskip=0pt
           \hbox{$<$}\hbox{$\sim$}}}}
\def\simgt{\mathrel{\lower2.5pt\vbox{\lineskip=0pt\baselineskip=0pt
           \hbox{$>$}\hbox{$\sim$}}}}

\newdimen\short
\def\adv{\global\advance\eq by1}
\def\set#1#2{\setbox#1=\hbox{#2}}
\def\nextlet#1{\global\advance\eq by-1\setbox
                \lett=\hbox{\rlap#1\phantom{a}}}

\newcount\eqncount
\eqncount=0
\def\equn{\global\advance\eqncount by1\eqno{(\the\eqncount)} }
\def\put#1{\global\edef#1{(\the\eqncount)}           }

\def\np{{\it Nucl. Phys.}}
\def\pl{{\it Phys. Lett.}}
\def\pr{{\it Phys. Rev.}}
\def\prl{{\it Phys. Rev. Lett.}}

\def\cpre{1}
\def\clep{2}
\def\cSS{3}
\def\cMS{4}
\def\cMSP{5}
\def\cGin{6}
\def\cSch{7}
\def\clwl{8}
\def\cGuf{9}
\def\cGuo{10}
\def\cmsu{11}
\def\cthr{12}
\def\cib{13}
\def\caekn{14}
\def\cAB{15}
\def\cDF{16}
\def\cFiq{17}
\def\cAnt{18}
\def\celn{19}
\def\cMoh{20}
\def\cbl{21}
\def\cRS{22}
\def\cPS{23}
\def\cCve{24}
\def\cGu3{25}

\magnification=1200
\hsize=6.0 truein
\vsize=8.5 truein
\baselineskip 14pt

\nopagenumbers

\rightline{hep-ph/9507219}
\rightline{IC/95/140}
\rightline{July 1995}
\vskip 1.0truecm
\centerline{ {\bf STRINGY UNIFICATION HELPS SEE-SAW MECHANISM}}
\vskip 1.0truecm
\centerline{{\bf Karim Benakli} \footnote{$^*$}{\it e-mail:
benakli@ictp.trieste.it}and {\bf Goran
Senjanovi{\'c}}\footnote{$^\dagger$}{\it e-mail:
goran@ictp.trieste.it}} \vskip .5truecm
\centerline{{\it International Centre for Theoretical Physics}}
\centerline{{\it 34014 Trieste, Italy}}
\vskip .5truecm

\vskip 2.5truecm
\centerline{\bf ABSTRACT}
\vskip .5truecm

In this paper we explore the possibility of intermediate scale
physics in the context of superstring models with {\it higher Kac-Moody
levels}, by focusing on left-right and Pati-Salam symmetries. We find
that the left-right scale may lie in the range $10^{10} - 10^{12}$
GeV which is favored by neutrino physics, while the Pati-Salam scale
is at most two or three orders of magnitude below the unification
scale $M_X$. We also show that the scale of $B-L$ breaking can be as
low as 1 TeV or so, providing protection against too rapid proton
decay in supersymmetry. Our results allow a natural value for the
scale  $M_X \sim 10^{18}$ GeV and the agreement with the experiment
requires the value of $\sin^2 {\theta_w}$ at $M_X$ to be in general
very different from the usually assumed $3/8$.

\hfill\break
\vfill\eject

\footline={\hss\tenrm\folio\hss}\pageno=1

\vskip 1.0cm
\centerline{\bf 1. Introduction}
\vskip 0.5cm

One of the main reasons to study supersymmetric theories is that
they could alleviate the problem of the gauge hierarchy. The
minimal supersymmetric extension of the standard model (MSSM) leads
to a remarkable prediction [\cpre]: the gauge couplings
$\alpha_3 $, $\alpha_2 $, $\alpha_1$ of $SU(3)_c$, $SU(2)_L$ and
$U(1)_Y$ respectively, are unified at a scale ${M_{\rm GUT}}\simeq
2\times 10^{16}{\rm GeV}$ [\clep] through the relation :
$$
\alpha_3 (M_{\rm GUT}) = \alpha_2 (M_{\rm GUT}) = {5\over 3} \alpha_1
(M_{\rm GUT}).\equn\put\one
$$

While very exciting, this result rests, however, on the hypothesis of
the ``Big Desert Scenario",which states that ``nothing" happens
between the scale of supersymmetry breaking ($\sim$ TeV) and the scale
of unification ($\sim M_{\rm GUT}$). At higher energies the
normalization factor ${5\over 3}$ of $U(1)_Y$ allows to embbedd
$SU(3)_c\times SU(2)_L\times U(1)_Y$ in a Grand Unified Theory (GUT)
group $SU(5), SO(10),\cdots$.

There are obvious reasons to look beyond this Big Desert
Scenario. The main one is the possibility of detecting some new
particles at future colliders. This needs
the intermediate scale physics to be of the order of TeV, which is
usually hard to obtain naturally. This is the case for
example of extra gauge bosons or more exotic such as an
extra-dimension.

Another important motivation for intermediate scales is the
question of neutrino masses. In the MSSM neutrino masses are made to
vanish  by hand, through the requirement of the absence of their right
handed partners. If the latter are present, the small values of
neutrino masses could naturally come from the see-saw mechanism
[\cSS,\cMS]. If right handed neutrinos get majorana masses at some
scale $M_{\nu_R}$, one expects a mass matrix of the form:

$${\left(\matrix{0 & m_D\cr m_D & M_{\nu_R}\cr}\right)}
\equn\put\two
$$

where $m_D$ are the Dirac masses of the neutrino. If $m_D << M_{\nu_R}$
then the masses of the right and left handed neutrinos are $M_{\nu_R}$
and ${m^2_D \over M_{\nu_R}} << M_{\nu_R}$, respectively. For the
see-saw mechanism to give us the neutrinos masses,
the large scale $M_{\nu_R}$ should be predicted. Normally, one
associates this scale with the breaking of some (gauge) symmetry. For
example, this can be naturally implemented if at some intermediate
scale $M_I$, the symmetry is enhanced to a left-right group
[\cMSP]\footnote{$^1$}{ Hereafter, for us $M_I$ will denote any
intermediate scale: $M_W << M_I << M_X$ where $M_X$ is the unification
scale.}.

In the standard model, left handed quarks and leptons are doublets of
$SU(2)_L$, while their right partners are singlets. In left-right
models, the explicit violation of Parity is replaced with
a spontaneous one, rendering our world more symmetric. In fact right
handed quarks and leptons (and thus neutrinos too) appear now also in
doublet representations, but under another gauge group $SU(2)_R$. To
relate these models to the MSSM, one has to introduce two scales:
the scale of parity breaking $M_{Parity}$ and the scale of $SU(2)_R$
breaking $M_R$, with obviously $M_R \le M_{Parity}$. It then natural
to relate these scales as $M_{\nu_R} \sim M_R$ and $M_{Parity} \sim
M_R$ or $M_{Parity} \sim  M_X$.

The simplest realization of this idea needs the introduction of
a Higgs triplet of $SU(2)_R$, usually denoted $\Delta_R$ that gets a
vacuum expectation value $M_R$. Anomaly cancellation imposes the
presence of another field $\bar {\Delta}_R$. If one wants to
have $M_{Parity}$ of the order of $M_R$, triplets $\Delta_L + \bar
{\Delta_L}$ of $SU(2)_L$  must also be introduced. A natural question
to ask is what happens to the gauge coupling unification when these
new states are present.

In all previous analysis, the unification of
these models was supposed to appear inside an $SO(10)$ gauge group,
and the above triplets arise from ${\bf 126}$ representations. It is
easy to see that unification constraints then imply that $M_R \sim
M_{GUT}$, and thus preventing the existence of such an intermediate
scale. Thus  these models were studied without refering to
unification. Moreover, new {\it ad-hoc} and more complicated states
are often introduced with the only purpose to lead to $SO(10)$
unification.

In this paper, we want to study the possibility of string unification
with arbitrary Kac-Moody levels for these models. In fact, these simple
models with triplets of $SU(2)_R$ are the first phenomenologically
interesting ones, well motivated by the see-saw mechanism, that would
{\it need stringy unification}. It is well known that this type of
unification does not require the existence of a GUT group (see below).
We will see below that, in contrast to all previous analysis, values of
$\sin^2 {\theta_w}$ at the unification scale very different from $3/8$
could be in agreement with the experimental data. We will also use this
opportunity to make some useful comments on the models.

In section 2 we review the problem of gauge couplings in string
theory. This will also allow us to define our notations and strategy.
In section 3 we will discuss the models and then determine some sets of
values for Kac-Moody levels leading to gauge couplings unification.
Comments of the results and conclusions are given in section 4.

\vskip 1.0cm
\centerline{\bf 2. Stringy unification of gauge couplings}
 \vskip 1.0cm
{ 2.1 STRING UNIFICATION AND MSSM}
\vskip 0.5cm

Superstring theory is considered today as a good
candidate for the unification of all known interactions, as we hope
that it contains a finite theory of quantum
gravity. It is thus always important to re-address the question of
unification of gauge couplings in its context.

In heterotic string theories, all the gauge (and the gravitational)
couplings are given by the expectation value of a particular field: the
dilaton. Moreover, the four dimensional space-time  gauge
symmetries are associated with an appropriate Kac-Moody algebra on the
two dimensional string worldsheet [\cGin]. To a Kac-Moody algebra
corresponds its level, a positive parameter $k_i$ (integer for
non-abelian groups), which determines the corresponding tree-level
gauge coupling constant  $\alpha_i$ in terms of the four-dimensional
string coupling $\alpha_{st}$, $\alpha_i = \alpha_{st} / k_i$. The
values of the levels also constrain the allowed unitary
representations present in the chiral massless spectrum [\cSch].

A natural question to ask then is: what are the values of the levels
$k_3$, $k_2$, $k_1$ associated with the standard model gauge groups
$SU(3)_c\, SU(2)_L$, and $U(1)_Y$ respectively? In most of the models
built up to now $k_3=k_2=1$, while $k_1 \ge {5 \over 3}$. These level
one constructions have the nice feature to explain the
presence of only singlet and fundamental chiral representations in
the standard model. However, they generally suffer from the presence of
fractionally charged particles in the massless spectrum at the
string level. Moreover, the most popular value of $k_1={5 \over 3}$
doesn't allow one to embbedd the MSSM in a GUT group because of the
absence of chiral adjoint representations.

In principle arbitrary higher Kac-Moody levels  $(k_3,k_2>1)$ are
allowed. However, the corresponding string models have been found
very difficult to build. They also allow the presence of
larger representations leading to phenomenological problems [\clwl].
For these reasons, they have been mainly disregarded. Recently, some
(still unsuccesfull) attempts have been made to build such theories,
with $k_3=k_2={3 \over 5}k_1=2$, trying to embbedd the standard
model is some GUT group and explain the {\it a priori arbitrary}
normalization $k_1$ of $U(1)_Y$ [\cGuf,\cGuo].

It is worth to notice that the (field theoretical) direct
unification of gauge couplings, which is the remarkable prediction
of the Big Desert Scenario, takes place at a scale
${M_{\rm GUT}}\simeq 2 \times 10^{16}{\rm GeV}$. However, in contrast to
the case in field theory,
the unification scale $M_{\rm SU}$ in string models can be predicted.
Within some large class of models, it was found to be $M_{\rm SU} \simeq
2\sqrt{\alpha_{st}} \times 10^{18}{\rm GeV}$, nearly two orders of
magnitude bigger than ${M_{\rm GUT}}$ [\cmsu]. Some ideas have been
presented to reconcile the two scales. They fall in two categories.

In the first category, one tries to push down $M_{\rm SU}$
toward ${M_{\rm GUT}}$ either by invoking large string threshold
corrections [\cthr], or by arguing on the possibility of a unified
evolution of the gauge couplings between these two scales. These
solutions have as good feature a small
ratio ${M_{\rm GUT}\over M_{\rm Planck}}$, which could be associated
with some explanations of the fermions mass spectrum, or with the
strength of the fluctuations in the COBE observations. Unfortunately,
the needed large thresholds do not seem to appear naturally.

In the second category, one tries to push ${M_{\rm GUT}}$ toward
$M_{\rm SU}$. This involves either the modification of the
hypercharge normalization (such as
\footnote{$^2$}{Such normalization doesn't appear in known level one
constructions.} $k_1 \simeq {4 \over 3}$) [\cib], or the presence of
some extra particles in some intermediate scale(s) [\caekn]. These
particles could be standard like or exotic fractionally charged ones
[\cAB]. These last possibilities are a clear abandon of the Big Desert
Scenario [\cDF]. In fact, they seem natural solutions as string
models usually contain more particles than the MSSM ones in the
massless spectrum at the string level, and some of them could be lying
somewhere between the TeV and the string scales.

\vskip 1.0cm
{ 2.2 STRINGY UNIFICATION WITH ONE INTERMEDIATE SCALE}
\vskip 0.5cm

Below, we would like to investigate the (next
to minimal) situation where some intermediate scale appears
corresponding to some symmetry breaking, with a minimal particle content
motivated by some phenomenological reasons. We ask about possible
existence of string unification for these models. This corresponds to
determine if there exist a set of levels $k_i$'s compatible with it.
Most of such models contain some large representations that need some
high level Kac-Moody algebras [\cFiq]. With our actual knowledge of
conformal field theories, building such compactifications is a
challenging problem. In realistic models, we would also have to
explain why and how only the wanted particles appear at the low and
intermediate scales. In particular, some representations (of smaller
conformal weight than the ones considered for example) could (probably
would) appear and spoil our analysis. In view of the above discussion,
we will not try to answer these problems, but being less ambitious,
we will constrain ourselves to the
analysis of the gauge coupling unification in such hypothetical
\footnote{$^3$}{ Neither the MSSM, nor its phenomenological viable
extensions (even some versions with extra chiral matter with
{\it appropriate spectrum} needed to raise the unification scale) have
been by now derived from strings.} cases.

We restrict our analysis to the one-loop unification of gauge
couplings in some particular supersymmetric models. We mainly consider
the possibility of one intermediate scale $M_I$ lying in the region
between the supersymmetry breaking scale $m_s$ and the unification
scale $M_X$.

Below $M_I$, the gauge group is the standard model $SU(3)_c\times
SU(2)_L\times U(1)_Y$ with corresponding Kac-Moody levels $k_3$,
$k_2$,and $k_1$ satisfying some constraints arising from the particle
content of the model below the string scale [\cSch]. More precisely, a
representation $(r_1, r_2,\cdots,r_n, q_1,\cdots,q_m)$ of
$SU(N_1)\times \cdots \times SU(N_n)\times U(1)_1 \times \cdots
\times U(1)_m$, of levels $k_{N_1} \cdots k_{N_n},k_{1_1}\cdots
k_{1_m}$, has a conformal weight: $$
h=\sum^n_{i=1} {C(r_i)\over {k_{N_i} + C(N_i)}}+\sum^m_{j=1}
{q^2_j\over k_{1_j}}
\equn\put\extra
$$
This state to be present in the string massless spectrum needs to
have $h \le 1$.

In the minimal case of MSSM content, $k_3$ and $k_2$
are positive integers while $k_1 \ge 1$. Notice that the case where all
the levels go to infinity correspond to the field theory limit as the
string scale goes to infinity and all the representations are allowed.
Unification is meaningless in this limit.

The associated effective couplings at the supersymmetry breaking scale
$m_s$ are given, at one loop by:

$$
{1 \over {\alpha_i (m_s)}} = {k_i \over \alpha_{st}}
+{b_i \over {2\pi}} {\ln ({M_I \over m_s})}+ {b'_i \over
{2\pi}} {\ln ({M_X \over M_I})}
 \equn\put\three
$$
where  $b_i$ and $b'_i$ are the one-loop beta-function coefficients
in the corresponding energy domains. They are given by:
$$
b_i= -3 C(G_i) + \sum_{{\rm reps} R_i} T(R_i).
\equn\put\four
$$
where the quadratic Casimir $C(G_i)$ of the group $G_i$ equals
$N$ for $SU(N)$ and $N-2$ for $SO(N)$. The index $T(R_i)$ of the
matter representation $R_i$ is equal to ${1 \over 2}$ for chiral
supermultiplets in the fundamental representation of $SU(N)$,  while
it is given by the sum of the squares of charges in the case of
$U(1)$.

The perturbative unification at the scale $M_X$ imposes a strong
constraint $\alpha_{st} / k_i <1$, which rules out the possibility of
a low $M_I$ scale in most of our models.

We would like to have some reasonable constraints on the
allowed values of $k_i$s. We first notice that the string unification
scale is predicted to be of the order of:

$$
M_{\rm SU} \simeq 2\sqrt{\alpha_{st}} \times 10^{18}{\rm GeV}
= 2\sqrt{\alpha_i k_i} \times 10^{18}{\rm GeV}
=2\sqrt{\alpha_3 k_3} \times 10^{18}{\rm GeV}
\equn\put\five
$$

We would like to keep $M_{\rm SU} \simeq 10^{18}{\rm GeV} <
M_{Planck}$. As in all our cases, $\alpha_3 \simgt 1/25$, this means
that $k_3$ should not exceed a number of order 100. A stronger
constraint could come if we assume the existence of an extra
non-abelian group with a smaller level $k < k_3$ then $k_3 \simlt
25k$. Moreover, another condition that could be imposed on the levels
is:  $$ {1 \over 3} k_3 + {1 \over 4} k_2 + {1 \over 4} k_1 = {\rm
integer} \equn\put\six
$$
which is required in order to avoid the appearance of fractionally
electrically charged particles in the massless spectrum [\cSch]. If
this condition is not satisfied, these undesired particles could
however still get masses of the order $M_X$ \footnote{$^4$}{ If they
are confined at very high scale by some extra gauge factor [\cAB,
\celn], or get a mass through one of the mechanisms discussed in
[\cFiq]}. Notice that what we mean by charge quantization is that all
color singlet states have a charge which is integer multiple of the
electron charge. In orbifold compactifications for instance, it has
been shown that a weaker charge quantization, where the elementary
charge is a fraction $1/N$ ($N \le 12$) of the electron charge, can be
imposed [\cAnt].

In {\three} we have absorbed the unknown string threshold
corrections ( usually denoted $\Delta_i$) in the definition of $M_X$
which can then be different from the computed value $M_{\rm SU}$.
While the {\it natural value of} $M_X$ {\it is} $\simeq 10^{18}{\rm
GeV}$, for practical computations, we allow it to take values between
$10^{16} {\rm GeV}$ and (more natural value) $10^{18}{\rm GeV}$. The
former value has the advantage of introducing naturally a small ratio
in the theory and thus it is often considered as a good value in the
literature.

\vskip .3truecm

In addition to $M_X$, our other inputs are of two kind:

-as experimental inputs within our one-loop approximations, the values
of the  strong , electromagnetic coupling constants at $m_Z$,
$\alpha_s \equiv \alpha_3$, $\alpha_{em} = {{\alpha_1 \alpha_2}\over
{\alpha_1 +\alpha_2}} $ respectively, and the weak angle $s \equiv
\sin^2{\theta}_w = {{\alpha_1 }\over {\alpha_1 +\alpha_2}}$ will be
taken in the range:
$$
\alpha_{em}= 1/128\, \qquad  0.230 \simlt s  \simlt 0.233\, \qquad
{\rm and}\quad  0.11 \simlt \alpha_s \simlt 0.13
\equn\put\seven
$$

-as theoretical inputs, we take
$m_S = m_Z$, which is usually a good approximation at one loop. Below
$M_I$, we will make the assumption that the massless spectrum is the
one of the MSSM with three generations ($n_g =3$) and two Higgs
doublets ($n_H =2$). The coefficients $b_i$ take the values:
$$
b_3=-9+2 n_g =-3 \; \qquad b_2=-6+2 n_g + n_H /2 =1\; \qquad
b_1=(10/3) n_g + n_H /2 =11
\equn\put\eight
$$

Our strategy is to solve {\three} in order to get the
ratios $k_1 / k_2$ and $k_2 / k_3$ as function of $\ln ({M_X \over
M_I})$:

$$
{k_1 \over k_2}= { {1-s- {\alpha_{em} \over {2 \pi}} \left[b_1
\  {\ln ({M_I \over m_Z})}+ (b'_1 - b_1) \  {\ln ({M_X \over
M_I})} \right]} \over  {s- {\alpha_{em} \over {2 \pi}} \left[b_2 \
{\ln ({M_I \over m_Z})} + (b'_2 - b_2) \  {\ln ({M_X \over M_I})}
\right]}}
\equn\put\nine
$$

$$
{k_2 \over k_3}= {  {{s \over \alpha_{em}}- {1 \over {2
\pi}} \left[b_2 \  {\ln ({M_I \over m_Z})} + (b'_2 - b_2) \  {\ln
({M_X \over M_I})} \right]} \over  {{1 \over \alpha_s}- { 1 \over {2
\pi}} \left[b_3 \  {\ln ({M_I \over m_Z})}+ (b'_3 - b_3) \  {\ln ({M_X
\over M_I})} \right]}}
\equn\put\ten
$$

 By plotting these functions as well as the values of $\alpha_{st} /
k_i$ which have to remain small, one can read the allowed intermediate
scale and the corresponding ratios of levels.

\vskip 1.0cm
\centerline{\bf 3. Models with intermediate mass scales}

\vskip 1.0cm
{ 3.1 THE MODELS}
\vskip 0.5cm

We want to focus on a single intermediate scale, although we shall
also discuss a case with two such scales. Our analysis
continue the analysis of [\cFiq] and it is parallel to
the ones for $SO(10)$ unification. There are four possible rank five
gauge groups at the intermediate scale, with their respective levels:

\vskip .3truecm

a)$SU(3)_c \times SU(2)_L \times U(1)_R \times
U(1)_{B-L}$ with levels $k_3$, $k_2$, $k_R$ and $k_{B-L}$. This model
has two nice features, if the scale $M_I$ lies in the TeV region. On
one side it predicts the observation of a new vector boson at future
colliders, and on the other hand the $B-L$ gauge
symmetry forbidds the appearance of (nonrenormalizable) operators
leading to fast proton decay.

\vskip .3truecm

b)$SU(3)_c \times SU(2)_L \times SU(2)_R \times
U(1)_{B-L}$ with levels $k_3$, $k_2$, $k_{2R}$ and $k_{B-L}$. In
addition to the aesthetic beauty of parity as a symmetry [\cMSP], these
models can explain small neutrino masses through the see-saw mechanism
[\cMS]. Two cases are a priori allowed and will be analysed below. The
first case corresponds to the direct breaking to the standard model at
the scale $M_I$. If this happens at the TeV scale, then $B-L$
symmetry protects the proton from fast decay [\cbl]. The model predicts
then the observation of extra neutral and charged gauge bosons.
However, in the natural approximation that the neutrino Dirac masses
are of the same order as the corresponding charged lepton masses, the
see-saw mechanism leads to a spectrum of too heavy left handed
neutrino masses overclosing the universe.

Namely, with $m_D \simeq m_l$ and $M_{\nu_R} \simeq 1$ TeV, we
predict
$$
m_{\nu_e}\simeq 1 {\rm eV} \qquad m_{\nu_\mu} \simeq 10 {\rm
keV} \qquad m_{\nu_\tau} \simeq 1-10 {\rm MeV}
\equn\put\neutrmass
$$

Now, $\nu_\tau$ can in principle decay through the weak currents:
$\nu_\tau \rightarrow e^+ e^- \nu_e$, by assuming the CKM-like matrix
in the leptonic sector. The case of $\nu_\mu$ is more problematic and
it requires the presence of $\Delta_L$, a left-handed analog of
$\Delta_R$ [\cRS].

On the other hand, if all $\nu$'s are lighter than 10-100 eV, then we
have no problem with the overclosure of the universe, and $\Delta_L$'s
are not necessarily present. In this case we have a constraint
$M_{\nu_R} \ge 10^8$ GeV (assuming the relation $m_D=m_l$ as in the
above).

If we wish to have the MSW explanation of the solar neutrino puzzle
(through $\nu_e-\nu_\mu$ oscillations), a prefered value for the
intermediate scale becomes $M_I\simeq 10^{10}$ GeV, with
$$
m_{\nu_e}\simeq 10^{-7} {\rm eV} \qquad m_{\nu_\mu} \simeq 10^{-3}
{\rm eV} \qquad m_{\nu_\tau} \simeq 1 {\rm eV}
\equn\put\neutrmass
$$
in which case $\nu_\tau$ can play a role of dark matter (or some
fraction of it).  Due to the uncertainties in $m_D$, we quote this as
$M_I \simeq 10^8-10^{12}$ GeV. In this case, $B-L$ does not protect the
proton from decaying too fast.

A spontaneous breaking of R-parity giving a vev to the
sneutrino $<\tilde{\nu}_R>$ could lead to proton decay through
dimension four operator in the superpotential. One then could
introduce a discrete symmetry or look for models where such operators
are forbidden by some string selection rules.

 Another possibility to get rid of this problem is to break the
group in two steps: first to $SU(3)_c \times SU(2)_L \times U(1)_R
\times U(1)_{B-L}$ and then to break it again at
the TeV scale to the standard model group, protecting the
proton from decaying too fast. We will investigate a minimal
version of this scenario too.

\vskip .3truecm
c)$SU(4)_c \times SU(2)_L \times U(1)_R $ with levels $k_4$,
$k_2$ and $k_R$. This partially unified model provides a symmetry
between quarks and leptons.

\vskip .3truecm
d)$SU(4)_c \times SU(2)_L \times SU(2)_R$ with levels $k_4$,
$k_2$ and $k_{2R}$. This Pati-Salam partial unification is the
minimal unification based on simple group of the standard group [\cPS]
and offers both left-right symmetry and quark-lepton unification.

\vskip 1.0cm
{ 3.2 UNIFICATION CONSTRAINTS}
\vskip 0.5cm

a) $SU(3)_c \times SU(2)_L \times U(1)_R \times U(1)_{B-L}$

The minimal content of matter is just the
MSSM spectrum plus three chiral superfields with quantum numbers $\nu_R
=(1,1,-1/2,1)$ under $SU(3)_c \times SU(2)_L \times SU(2)_R \times
U(1)_{B-L}$ group  , which can then be identified as  right handed
neutrinos. The condition imposed by the presence of these states on
the Kac-Moody levels is:   $$
{1 \over {4 k_R}} + {1 \over k_{B-L}} \le 1
\equn\put\eleven
$$
The level $k_1$ given by:
$$
k_1=k_R+{k_{B-L} \over 4 }
\equn\put\twelve
$$
can take the standard value $k_1=5/3$. When one of the new fields
$<\tilde{\nu}_R>$ gets a vev, it breaks one combination $U(1)'$ of
$U(1)_R \times U(1)_{B-L}$ leading to an extra $Z'$ massive vector
boson at the scale $M_I$, while it leaves the hypercharge $U(1)_Y$
with $Y=Q_R + (B-L)/2$ unbroken, where $Q_R$ is the generator of
$U(1)_R$. How would the neutrinos get a mass in this case? One
possibility is then the mechanism suggested in [\cMoh]. A see-saw
mechanism is obtained through the mass matrix between the left and
right neutrino and the gaugino partner of $Z'$. This however gives a
mass only to the neutrino whose partner got a vev. The other neutrinos
presumably get masses through some non-renormalizable operators.

 Another possibility is to explain neutrino masses by
the usual see-saw mechanism. The gauge symmetry breaking is
achieved by giving vev to some extra state with gauge numbers
$(1,1,1,-2)$ (one also introduces a $(1,1,-1,2)$ representation to
cancel the $U(1)$ anomalies). Then one has the condition:
$$
{1 \over  k_R} + {4 \over  k_{B-L}} \le 1
\equn\put\thirteen
$$
which implies that $k_R >1$ and $k_{B-L} > 4$ so $k_1 > 2$. As the new
state couple equally to all the neutrinos, they generate
small masses to all of them through the see-saw mechanism.
An important question to raise is what are the expected values for
the Dirac masses. In the general case we are considering, it is not
possible to make a model independent statement. As the right handed
neutrinos and electrons are a priori independent, and could come
from different sectors of the string compactification, having different
moduli dependence, the relative Yukawa couplings could be
very different. Having smaller values for $m_D$ corresponding to
$\nu_{\mu}$ would allow a low scale $M_I$.

In any case the gauge couplings of $SU(3)_c\times SU(2)_L\times
U(1)_Y$ are not affected and evolve in the same way as in the MSSM,
with the new appropriate normalizations $k_i$. The scale $M_I$ is
only constrained by collider experiments, and the corresponding gauge
coupling of $U(1)'$ can be now computed, because at $M_X$, it is
equal to the one of $U(1)_Y$ ($k' = k_R+{k_{B-L} \over 4 } = k_1$),
and the associated beta-function coefficient is known: $b_1$ or
$b_1+2=13$ in the first and second examples described
above respectively. At low energies, the corresponding coupling is
equal or smaller than the hypercharge one.

\vskip 0.5truecm

b) $SU(3)_c \times SU(2)_L \times SU(2)_R \times U(1)_{B-L}$

This case is more interesting because it is more
restricting. The minimal matter content of the model is $n'_g$
generations of matter representations $Q=(3,2,1,1/3)$,
$Q^c=(3,1,2,-1/3)$, $L=(1,2,1,-1)$, and $L^c=(1,2,1,1)$ which
correspond to the quarks and leptons. There is also a Higgs sector
consisting in $n_{22}$ bidoublets $(1,2,2,0)$, $n_{\Delta_L}$ pairs of
$SU(2)_L$ triplets $\{ \Delta_L =(1,3,1,2) + \bar{\Delta}_L
=(1,3,1,-2) \}$ as well as $n_{\Delta_R}$ pairs of $SU(2)_R$ triplets
$\{ \Delta_R =(1,1,3,-2) + \bar{\Delta}_R =(1,1,3,2) \}$ and a possible
parity odd singlet to make $n_{\Delta_L}=0$ [\cCve]. The $\Delta_R$
field, when it gets a vev, breaking $SU(2)_R \times U(1)_{B-L}$ to
$U(1)_Y$, gives a Majorana mass to the right handed neutrino.

The coefficients of the beta-functions of $SU(3)_c$, $SU(2)_L$, and
$U(1)_Y$ above $M_I$ are:
$$
\eqalign {b'_3 &=-9+2 n'_g + n_3 \; \qquad b_2=-6+2 n'_g + n_{22}+ 4
n_{\Delta_L} + n_2 \; \cr
b'_1 &=-6 + (10/3) n'_g +n_{22}+ 6 n_{\Delta_L} + 10 n_{\Delta_R}+n_1
\cr}
\equn\put\fourteen
$$

respectively. In all of the discussions below, the numbers $n_1$, $n_2$
and $n_3$ will parametrize the (unknown) contributions of extra
particles that could appear at this scale. Unless explicitly stated
otherwise, we take  $n'_g=3$, $n_1 = n_2 = n_3 =0$. The normalization
of $U(1)_Y$ is given by:
$$
k_1 = k_{2R} + {k_{B-L} \over 4 }
\equn\put\fifteen
$$
The minimal values of levels are $k_{2R} \ge 2$, $k_{B-L} \ge 8$
(hence $k_1 \ge 4$), $k_2 \ge 2$ if $n_{\Delta_L} \neq 0$. One can
define two possible left-right symmetries: one with equal couplings
for $SU(2)_L$ and $SU(2)_R$ implying $k_2=k_{2R}$ and $k_1 \ge k_2 +2$
and the other with different couplings $k_2 \neq k_{2R}$.  In the
more symmetric case, one has the constraint ${k_1 \over k_2} \ge 1$,
the equality corresponding to $k_2 \rightarrow \infty$. This constraint
is too restrictive and it leads usually to large intermediate scales.
We will relax this constraint and allow for $k_2 \neq k_{2R}$. Hence,
${k_1 \over k_2} \le 1$ is allowed, and the limits will come from the
perturbative unification limit on $\alpha_i$ at the unification scale.

We have made the analysis for different particles content and
different unification scales, and we present our results in tables 1,
2, 3 and the figures. In particular we studied the cases:

i) $n_{22}=1$, $n_{\Delta_R}=1$, and $n_{\Delta_L}=0$. The
results are plotted in figures 1 and 2.  This model could be made
more symmetric if one explains $n_{\Delta_L}=0$ through the introduction =
of
some parity odd singlet [\cCve]. While from the figures one can
read that the intermediate scale as low as few TeV is allowed by the
runnings of the couplings, the neutrino spectrum forbidds it. In
fact as discussed above, one gets too heavy $\nu_{\mu}$ which is stable
because of the absence of $\Delta_L$, thus overclosing the universe.
We then have a constraint $M_I \simgt 10^7$ GeV as discussed above.

If one insists on the equality of left and right couplings, then for
$k_2 \simeq 10$, we get $M_I \simgt 10^{14.5}, 10^{11}, 10^{8} $ GeV
for $M_X= 10^{18}, 10^{17}$ and $10^{16}$ GeV, respectively. A possible
set, for $M_X= 10^{18} {\rm GeV}$, is $k_3=11$, $k_2=k_{2R}=10$ and
$k_{B-L}=8$.

\vskip 0.5truecm

ii) $n_{22}=2$, $n_{\Delta_R}=1$, $n_{\Delta_L}=0$. This case is
quite similar to  the first one. As we are interested in large
$M_I$ ( $\simeq 10^{12}$ GeV) the ratio $k_1 / k_2$ is not
sensible to the number of bidoublets as shown in figure 3. The
addition of the extra bidoublet is helpful to generate a correct
Cabibbo angle. We present the corresponding results in figures 4 and
5.

\vskip 0.5truecm

iii) $n_{22}=1$, $n_{\Delta_R}=1$, $n_{\Delta_L}=1$. This is the
minimal fully left-right symmetric model allowing for see-saw
``explanation" of the neutrino masses. For this model we present our
results in figures 6 and 7. From these figures, we can see that now
the scale $M_I$ can not be as low as TeV, otherwise the hypercharge
coupling will blow up before the scale $M_X$. In order that this does
not happen, we get the lower value of
$M_I \simeq 10^{10} {\rm GeV}$.

\vskip 0.5truecm

iv) $n_{22}=2$, $n_{\Delta_R}=1$, $n_{\Delta_L}=1$. This is the most
popular model. As the perturbative condition of the couplings doesn't
allow small $M_I$, the contribution of the second bidoublet is small.
The results are displayed in figures 8 and 9. The end result is again
that the region $10^{10}-10^{12}{\rm GeV} $ is perfectly OK.

Thus one can have a realistic left-right model in the context of
strings with the MSW mechanism and $\nu_\tau$ as (some portion of)
the dark matter of the universe.

The tables 3 and 4 give some possible values of the Kac-Moody levels
consistent with $M_I$ of the order of $10^{10}-10^{12}{\rm GeV} $ or
electric charge quantization respectively.

\vskip 0.5truecm

v) We would like to investigate the possibility (from the gauge
couplings unification point of view) to have a low lying $B-L$
breaking scale around the TeV, protecting the proton from decaying; and
a left-right breaking scale at an intermediate scale. This two
intermediate  breaking scales could be achieved for example with the
following set of representations (in addition to the quarks,
leptons and Higgses):

\vskip 0.3truecm

Between $M_I$ and $M_X$ we have $n_{\Delta_L}$ pairs of $SU(2)_L$
triplets $\{\Delta_L =(1,3,1,2) + \bar{\Delta}_L =(1,3,1,-2) \}$,  as
well as $n_{\Delta_R}$ pairs of $SU(2)_R$ triplets $\{\Delta_R
=(1,1,3,-2)+\bar{\Delta}_R =(1,1,3,2)\}$ and new triplets $n_{T L}$
pairs of $SU(2)_L$ triplets $\{T_L =(1,3,1,0)+\bar{T}_L =(1,3,1,0)\}$
and corresponding $n_{T R}$ pairs of $SU(2)_R$ triplets $\{T_R
=(1,1,3,0)+\bar{T}_R =(1,1,3,0)\}$\footnote{$^5$}{The doubling of
states is dictated by the vanishing of the Fayet-Iliopoulos term when
one of these fields get a vev}. At $M_I$ the neutral component of $T_R$
gets a vev and it breaks $SU(2)_R$ to $U(1)_R$. The beta-functions
coefficients are given by:

$$
\eqalign {b'_3 &=-9+2 n'_g+n_3  \; \qquad b_2=-6+2 n'_g + n_{22}+ 4
n_{\Delta_L} + 4 n_{T_L}+n_2 \; \cr
b'_1 &=-6 + (10/3) n'_g +n_{22}+ 6 n_{\Delta_L} + 10 n_{\Delta_R}+4
n_{T_R}+n_1 \cr}
\equn\put\sixteen
$$
For our analysis we take  $n'_g=3$, $n_1 = n_2 = n_3 =0$, $n_{22}=
=1$ or
2, $n_{\Delta_R}=1$, $n_{\Delta_L}=1$ and $n_{T_L}=n_{T_R}=1$.

 At
$M_I$ where the new field $T_R$ is supposed to get a vev, some of
the above particles become massive and decouple from
the running of coupling constants below $M_I$.

 In the absence of a
complete analysis of the minimization of the full superpotential and
lacking a known extended survival principal for these models,
we take the minimal phenomenologically viable
spectrum to provide the light particles. For instance, we assume that
below $M_I$, in addition to the particle content of the standard
model, only the neutral component $\Delta^0_R$ of $\Delta_R$, and
all of $\Delta_L$ remain massless\footnote{$^6$}{If $\bar {\Delta}_L$
also remains massless, then $M_I \sim M_X$.}. Possible values of
intermediate scales and corresponding Kac-Moody levels are displayed
in table 3. It is worth noticing that in the case of $M_X \simeq
10^{16}$ GeV, one has the possibility of $k_1/k_2$ of order $5/3$ for
an $M_I \simeq 10^{10}$ GeV. This allows the hope to embedd it in an
$SO(10)$ GUT. However, to cure $k_3/k_2 >1$ one has to introduce a
large extra contribution $n_3$ for $b_3$, for example $n_3=10$ for
$\alpha_s \simeq 0.13$ which thus is extremely sensible to the exact
value of $\alpha_s$. In short, it is possible to keep $B-L$
symmetry in the TeV region with $M_I \simeq 10^{10}-10^{14}{\rm GeV}
$

\vskip 0.5truecm

c) $SU(4)_c \times SU(2)_L \times U(1)_R $

In this case, the
quarks and leptons are unified in the same representations of $SU(4)_c
\times SU(2)_L \times U(1)_R$ leading to predict the existence of the
right neutrinos and the size of their expected Dirac masses. In
addition to the quarks and leptons in the $n'_g$
representations:  $$
(4,2,1) + (4,1,-1/2) + (4,1,1/2)
\equn\put\seventeen
$$
and $n_h$ electroweak Higgs:
$$
(1,2,-1/2) + (1,2,1/2)
\equn\put\eighteen
$$

we have $n_{10}$ pairs of representations $(10,1,-1)$ and $(\bar
{10},1,1)$ necessary to break $SU(4)_c \times SU(2)_L \times U(1)_R$
to the $SU(3)_c \times SU(2)_L \times U(1)_Y$.

The standard model levels are given by:

$$ k_1= {2 \over 3} k_4 + k_R, \qquad \qquad \qquad k_3= k_4
\equn\put\nineteen
$$

The associated beta-functions coefficients above the intermediate
scale are:
$$
\eqalign {b'_3 &=-12 +2 n'_g +6 n_{10}+ n_3  \; \qquad b_2=-6+2 n'_g +
n_h + n_2 \; \cr
b'_1 &=-8 + (10/3) n'_g +n_h+ 24 n_{10}+n_1 \cr}
\equn\put\twenty
$$

We made the analysis for $n'_g=3$, $n_1 = n_2 = n_3 =0$, $n_h=1$,
$n_{10}=1$. We display in figures 10 and 11 the results for $M_X
\simeq 10^{18}$ GeV. In table 4, we give some values of levels
allowing intermediate scale and electric charge quantization.

\vskip 0.5truecm

d) Pati-Salam group $SU(4)_c \times SU(2)_L \times SU(2)_R$

The Pati-Salam group needs large representations to get broken to the
standard model one. The contribution of these particles to the running
of $U(1)_Y$ would make this coupling blow up very close to $M_I$. In
fact $M_I$ is typically two to three orders of magnitude below $M_X$
when the latter goes from $10^{18}$ to $10^{16}$ GeV.
Hence it is not very appealing as an intermediate scale, since only
for a low value of $M_X$ it becomes interesting for neutrino physics.

\vskip 1.0cm
\centerline{\bf 4. Discussion and Conclusion}
\vskip 0.5cm

Our results are mainly qualitative, and need some comments.
The maximal values of ${k_1 \over k_2}$ are obtained
in the absence of an intermediate scale. They vary between $1.4$ for
$M_X\simeq 10^{18}$ GeV  to $\simeq 5/3$ for $M_X\simeq 10^{16}$ GeV,
confirming previous analysis for the MSSM. In the minimal models, we
found that interesting values of the scale $M_I$ are allowed.
 Furthermore, since
renormalization group equations usually make $U(1)_Y$ coupling
increase with energy faster than $SU(2)_L$ one, $\sin^2 \theta_w$ at
$M_X$ is typically bigger than $3/8$.
Also, the necessary levels are often large, especially when one
imposes the condition {\six} for charge quantization. The cases
of large values of $k_3$ could be improved if one
allows the presence of an octet of $SU(3)$ for example.  An increase
of precision of our analysis, by improving the uncertainty on
$\alpha_s$ and by a serious two loops analysis with the thresholds
taken into account, could constraint the allowed values of the
unification scale or intermediate scale for some of these simple
models. For instance, in model (b i) $k_3/k_2$ is constant and if it
takes a value like 1.02, we can not associate it with two small
integers $k_3$ and $k_2$ ($k_3 \simlt 25$).

We would like to comment about the actual
status of string model building of higher Kac-Moody levels models.
The actual known trick to get such models is to notice that if
you take the ``diagonal" part of the product of n factors of the
same group $G$ at level one, you generate a gauge group $G$ of level
$n$. This method cannot generate the minimal spectrum considered
above. For instance, a triplet of $SU(2)$ comes from the product of
two doublets and thus is always accompanied by a singlet. We have
considered the effect of the additional singlets carrying $B-L$
charges on our analysis. We found a notable effect. As an example, the
deviation for model (b iv) is plotted in figures 12 and 13.

We would also like to compare with the case of usual
supersymmetric $SO(10)$ GUT. That case correspond to ${k_1 \over
k_2}={5 \over 3}$ and $k_3 = k_2$. From the analysis of our plots, it
is obvious that there cannot be an intermediate scale with the
content considered in this paper. The introduction of an intermediate
scale would there necessitate an {\it ad hoc} introduction of a set of
particles necessary for {\it fitting} the experimental data with a
possible unification in an $SO(10)$. From this point of view, the
beauty of the prediction of the MSSM for a unification of couplings is
totally lost.

Two kind of string constructions of these GUTs have been
investigated by now, using the method sketched above. The first uses
fermionic constructions [\cGu3]. In this case, it has been obtained
that possible Kac-Moody levels (normalization of any {\it non-abelian}
groups) are $k=1,2,4,8$. This allows only for ratios $k_3/k_2$ equal
to ${1 \over 8}, {1 \over 4}, {1 \over 2}, 1, 2, 4, 8$ which is not
satisfied in most of our cases.

The other analysis uses orbifold constructions [\cGuo] and it
indicates that some extra states, the part of the adjoint that are not
eaten by the Higgs mechanism, must remain massless at the GUT scale.
They are in representations:

$$ (8,1,0) + (6,1,-2/3) + (6,1,2/3) + (1,3,0) + (1,3,1)+ (1,3,-1)
\equn\put\twentyone$$

or

$$ (8,1,0) + (1,3,0) + (1,1,0)+ (1,1,1)+ (1,1,-1)
\equn\put\twentytwo
$$

They should then be lying somewhere, between the GUT and the
electroweak scales. We found that the
addition of these particles with a unique common mass (as
one intermediate scale) always destroy the GUT unification prediction.

In conclusion, we have presented an analysis of possible string
unification without GUT for simple and motivated extensions of the
minimal supersymmetric standard model which, in contrast to the MSSM,
necessarily need a departure from the more attractive level one string
constructions.

We  have focused on left-right and Pati-Salam symmetries and our
analysis shows that $M_R$ can be as low as $10^5$ GeV or so, and
furthermore the value $M_R \simeq 10^{10}-10^{12} {\rm GeV}$
interesting for neutrino physics is perfectly consistent with
unification constraints. In the case of two step breaking, we can
have $M_R$ as $M_I$ in the range $10^{14}-10^{11} {\rm GeV}$, while
allowing the $B-L$ gauge symmetry to remain unbroken all the way down
to TeV. Among other effects, this can save the proton from
decaying too fast which is a generic problem in supersymmetric
theories.

For the Pati-Salam scale, $M_{PS}$, we find that it has to be quite
large, some two to three orders of magnitude below $M_X$. Only if we
push $M_X$ down to $10^{16}$ GeV (not so appealing to us), we can
keep $M_{PS}$ as low as $10^{13} {\rm GeV}$ to provide an interesting
intermediate scale.

As optimistic point of view would be that these models can have
interesting intermediate scales, allowing a correct string unification
scale and charge quantization. A pessimistic point of view is that
the predicted values of the associated Kac-Moody levels and our poor
knowledge of how to build such theories make it hopeless to construct
these simple models in the near future, and they can only be studied
from the effective field theory point of view keeping in mind that the
gauge coupling unification could be achieved through the presence of
only one tree level gauge coupling in heterotic string models.

\vskip 1.0cm
\noindent{\bf Acknowledgments}
\vskip.5cm
We thank I. Antoniadis and C. Aulakh for useful comments.

\vskip 1.5cm
\centerline{\bf REFERENCES}
\vskip 0.5cm

\parskip=-3 pt

\item{[{\cpre}]} S. Dimopoulos, S. Raby, and F. Wilczek, {\pr} {\bf D
24} (1981) 1681; N. Sakai, {\it Zeit. Phys.} {\bf C 11}
(1981) 153;L. E. Ib{\'a}{\~ n}ez and G. G. Ross, {\pl} {\bf 105B}
(1981) 439; M. B. Einhorn and D. R. T. Jones, {\np} {\bf B196} (1982)
475; W. J. Marciano and G. Senjanovi{\'c}, {\it Phys. Rev.} {\bf D25}
(1982) 3092 .\hfill\break

\item{[{\clep}]} For LEP data analysis see: J. Ellis, S. Kelley and
D.V. Nanopoulos, {\pl} {\bf 249B} (1990) 441 and {\bf 260B} (1991) 131;
U. Amaldi, W.de Boer and H. F\"urstenau, {\pl} {\bf 260B} (1991) 447;
P. Langacker and M. Luo, {\it Phys. Rev.} {\bf D44} (1991) 817. For
more recent discussion: P. Langacker and N. Polonsky, {\it Phys.
Rev.} {\bf D47} (1993) 4028 and {\it Phys. Rev.} {\bf D49} (1994)
1454; L. J. Hall and U. Sarid, {\prl} {\bf 70} (1993)
2673.\hfill\break

\item{[{\cSS}]} M. Gell-Mann, P. Ramond and R. Slansky, in {\it
Supergravity}, eds. P. van Niewenhuizen and D.Z. Freedman (North
Holland 1979); T. Yanagida, in Proceedings of {\it Workshop on Unified
Theory and Baryon number in the Universe}, eds. O. Sawada and A.
Sugamoto (KEK 1979).\hfill\break

\item{[{\cMS}]} R. N. Mohapatra and G. Senjanovi{\'c}, {\prl} {\bf 44}
(1980) 912 and {\pr} {\bf D 23} (1981) 165.\hfill\break

\item{[{\cMSP}]} J. C. Pati and A. Salam, {\pr} {\bf D 10} (1974)
275; R. N. Mohapatra and J. C. Pati, {\pr} {\bf D 11} (1975) 566; R.
N. Mohapatra and G. Senjanovi{\'c}, {\prl} {\bf 44}, (1980) 912 and
{\pr} {\bf D 23} (1981) 165 \hfill\break

\item{[{\cGin}]} P. Ginsparg, {\pl} {\bf 197B} (1987)
139.\hfill\break

\item{[{\cSch}]} A. Schellekens, {\pl} {\bf 237B} (1990) 363.\hfill\break

\item{[{\clwl}]} D. Lewellen, {\np} {\bf B337} (1990) 61.\hfill\break

\item{[{\cGuf}]} S. Chaudhuri, S.-w. Chung and J.D. Lykken,
preprint Fermilab-Pub-94/137-T, hep-ph/9405374; S. Chaudhuri, S.-w.
Chung, G. Hockney and J.D. Lykken, preprint hep-th/9409151;G. B.
Cleaver, preprint hep-th/9409096.\hfill\break

\item{[{\cGuo}]} G. Aldazabal, A. Font, L. E. Ib{\'a}{\~ n}ez and
A. M. Uranga preprint FTUAM-94-28 hep-th/9410206.\hfill\break

\item{[{\cmsu}]} V.S. Kaplunovsky, {\np} {\bf B307} (1988) 145 and {\it
Errata} STANFORD-ITP-838 preprint (1992).\hfill\break

\item{[{\cthr}]} L. Dixon, V. Kaplunovsky and J. Louis,
{\np} {\bf355} (1991) 649; J.P. Derendinger, S. Ferrara, C. Kounnas
and F. Zwirner, {\np} {\bf B372} (1992) 145; I. Antoniadis, K. Narain
and T. Taylor, {\pl}{\bf 267B} (1991) 37; I. Antoniadis, E. Gava and
K. Narain, {\pl} {\bf 283B} (1992) 209; {\np} {\bf B383} (1992) 93;
E. Kiritsis and C. Kounnas, preprint CERN-TH-7472-94  hep-th/9501020.
\hfill\break

\item{[{\cib}]} L. E. Ib{\'a}{\~ n}ez, {\pl} {\bf 318B} (1993)
73.\hfill\break

\item{[{\caekn}]} I. Antoniadis, J. Ellis, S. Kelley and D.V.
Nanopoulos, {\pl} {\bf 272B} (1991) 31.\hfill\break

\item{[{\cAB}]} I. Antoniadis and K.Benakli, {\pl} {\bf 295B} (1992)
219.\hfill\break

\item{[{\cDF}]} For a recent work in this direction, see for example:
K. Dienes and A. Faraggi, preprint IASSNS-HEP-95/12, hep-ph/9505018 and
IASSNS-HEP-94/113, hep-ph/9505046; S. P. Martin and P. Ramond {\pr}
{\bf D51} (1995) 6515. \hfill\break

\item{[{\cFiq}]} A. Font, L. E. Ib{\'a}{\~ n}ez and F. Quevedo,
{\np} {\bf B345} (1990) 389.\hfill\break

\item{[{\cAnt}]} I. Antoniadis, {\it Proceedings of Summer School in High
Energy Physics and Cosmology, Trieste} (1990) 677.\hfill\break

 \item{[{\celn}]} J. Ellis, J.L. Lopez and D.V. Nanopoulos, {\pl} {\bf
245B=
}
(1990) 375 and {\bf 247B} (1990) 257.\hfill\break
\item{[{\cMoh}]} R. N. Mohapatra, {\prl} {\bf 56} (1986)
561.\hfill\break

\item{[{\cbl}]}  R. N. Mohapatra, {\pr} {\bf D34} (1986)
3457;  A. Font, L. E. Ib{\'a}{\~ n}ez and F. Quevedo,
{\pl} {\bf B228} (1989) 79; S. P. Martin, {\pr} {\bf D46} (1992)
2769.\hfill\break

\item{[{\cRS}]} M. Roncadelli and G. Senjanovi{\'c}, {\pl} {\bf B107}
(1981) 59.\hfill\break

\item{[{\cPS}]} J. C. Pati and A. Salam, {\pr} {\bf D10} (1974)
275. \hfill\break

\item{[{\cCve}]} M. Cvetic, {\pl} {\bf 164B} (1985) 55.\hfill\break

\item{[{\cGu3}]} S. Chaudhuri, S.-w. Chung, G. Hockney and J.D. Lykken,
preprint hep-th/9501361. \hfill\break

\vfill\eject
\vbox {\tabskip=0pt \offinterlineskip\def\tablerule{\noalign{\hrule}}
\def\tv{\vrule height 20pt depth 5pt}\halign to 13cm {\tabskip=0pt
plus 20mm \tv\hfill\quad#\qquad\hfill &\tv\hfill\quad#\qquad\hfill
&\tv\hfill\quad# \qquad\hfill &\tv\hfill\quad#\quad\hfill
&\tv#\tabskip=0pt\cr\tablerule  ($n_{22}$, $n_{\Delta_L}$,
$n_{\Delta_R}$)  & $M_X$ in GeV & $M_I$ in GeV &
  $(k_{1}, k_{2}, k_{3})$
&\cr\tablerule   $(1, 0, 1)$&$10^{18}$&$10^{11.5}$&$(9,9,10)$
&\tabskip=0pt\cr $(1, 0, 1)$&$10^{17}$&$10^{11}$&$(6,5,5)$
&\tabskip=0pt\cr $(1, 0, 1)$&$10^{16}$&$10^{11}$&$(7,5,5)$
&\tabskip=0pt\cr $(2, 0, 1)$&$10^{18}$&$10^{11.5}$&$(10,10,12)$
&\tabskip=0pt\cr $(2, 0, 1)$&$10^{17}$&$10^{12}$&$(25/2,10,11)$
&\tabskip=0pt\cr $(2, 0, 1)$&$10^{16}$&$10^{12}$&$(6,4,4)$
&\tabskip=0pt\cr $(1, 1, 1)$&$10^{18}$&$10^{12}$&$(6,8,14)$
&\tabskip=0pt\cr $(1, 1, 1)$&$10^{17}$&$10^{11}$&$(6,6,10)$
&\tabskip=0pt\cr $(1, 1, 1)$&$10^{16}$&$10^{10}$&$(5,4,6)$
&\tabskip=0pt\cr $(2, 1, 1)$&$10^{18}$&$10^{12}$&$(9/2,6,12)$
&\tabskip=0pt\cr $(2, 1, 1)$&$10^{17}$&$10^{11}$&$(5,5,9)$
&\tabskip=0pt\cr $(2, 1, 1)$&$10^{16}$&$10^{10}$&$(5,4,7)$
 &\tabskip=0pt\cr\tablerule}}

\vskip 1.0truecm
\centerline{\bf Table 1.}
\vskip 0.5truecm
Examples of values of Kac-Moody levels $(k_{3}, k_{2}, k_{1})$ of
$SU(3)_c\times SU(2)_L\times U(1)_Y$ allowing for $SU(3)_c \times
SU(2)_L \times SU(2)_R \times U(1)_{B-L}$ at an intermediate scale of
order $M_I \simeq 10^{12}$ GeV.

\vfill\eject
\vbox {\tabskip=0pt \offinterlineskip\def\tablerule{\noalign{\hrule}}
\def\tv{\vrule height 20pt depth 5pt}\halign to 13cm {\tabskip=0pt
plus 20mm \tv\hfill\quad#\qquad\hfill &\tv\hfill\quad#\qquad\hfill
&\tv\hfill\quad# \qquad\hfill &\tv\hfill\quad#\quad\hfill
&\tv#\tabskip=0pt\cr\tablerule  ($n_{22}$, $n_{\Delta_L}$, $n_{\Delta
R}$)  & $M_X$ in GeV & $M_I$ in GeV &
  $(k_{1}, k_{2}, k_{3})$
&\cr\tablerule   $(1, 0, 1)$&$10^{18}$&$10^{7}$&$(22/3,10,11)$
&\tabskip=0pt\cr $(1, 0, 1)$&$10^{17}$&$10^{8}$&$(12,12,12)$
&\tabskip=0pt\cr $(1, 0, 1)$&$10^{16}$&$10^{8.5}$&$(19/3,5,5)$
&\tabskip=0pt\cr $(2, 0, 1)$&$10^{18}$&$10^{11.5}$&$(10,10,12)$
&\tabskip=0pt\cr $(2, 0, 1)$&$10^{17}$&$10^{12}$&$(44/5,8,9)$
&\tabskip=0pt\cr $(2, 0, 1)$&$10^{16}$&$10^{11}$&$(44/3,12,13)$
&\tabskip=0pt\cr $(1, 1, 1)$&$10^{18}$&$10^{13}$&$(32/3,12,19)$
&\tabskip=0pt\cr $(1, 1, 1)$&$10^{17}$&$10^{9}$&$(8,12,24)$
&\tabskip=0pt\cr $(1, 1, 1)$&$10^{16}$&$10^{8.5}$&$(12,12,21)$
&\tabskip=0pt\cr $(2, 1, 1)$&$10^{18}$&$10^{13}$&$(32/3,12,22)$
&\tabskip=0pt\cr $(2, 1, 1)$&$10^{17}$&$10^{11.5}$&$(40/3,12,20)$
&\tabskip=0pt\cr $(2, 1, 1)$&$10^{16}$&$10^{11.5}$&$(52/3,12,17)$
 &\tabskip=0pt\cr\tablerule}}

\vskip 1.0truecm
\centerline{\bf Table 2.}
\vskip 0.5truecm
Examples of values of Kac-Moody levels $(k_{3}, k_{2}, k_{1})$ of
$SU(3)_c\times SU(2)_L\times U(1)_Y$ allowing for $SU(3)_c \times
SU(2)_L \times SU(2)_R \times U(1)_{B-L}$ at an intermediate scale,
leading to charge quantization.

\vfill\eject
\vbox {\tabskip=0pt \offinterlineskip\def\tablerule{\noalign{\hrule}}
\def\tv{\vrule height 20pt depth 5pt}\halign to 13cm {\tabskip=0pt
plus 20mm \tv\hfill\quad#\qquad\hfill &\tv\hfill\quad#\qquad\hfill
&\tv\hfill\quad# \qquad\hfill
&\tv#\tabskip=0pt\cr\tablerule   $M_X$ in GeV & $M_I$ in GeV &
  $(k_{1}, k_{2}, k_{3})$
&\cr\tablerule   $10^{18}$&$10^{15}$&$(4,6,30)$
&\tabskip=0pt\cr $10^{17}$&$10^{13}$&$(4,4,24)$
&\tabskip=0pt\cr $10^{16}$&$10^{11}$&$(7,4,24)$
&\tabskip=0pt\cr\tablerule}}

\vskip 1.0truecm
\centerline{\bf Table 3.}
\vskip 0.5truecm

Examples of values of Kac-Moody levels $(k_{3}, k_{2}, k_{1})$ of
$SU(3)_c\times SU(2)_L\times U(1)_Y$ allowing for $SU(3)_c \times
SU(2)_L \times SU(2)_R \times U(1)_{B-L}$ at an intermediate scale
$M_I$ and $SU(3)_c \times SU(2)_L \times U(1)_R \times U(1)_{B-L}$ at
low energies.

\vskip 3.0truecm

\vbox {\tabskip=0pt \offinterlineskip\def\tablerule{\noalign{\hrule}}
\def\tv{\vrule height 20pt depth 5pt}\halign to 13cm {\tabskip=0pt
plus 20mm \tv\hfill\quad#\qquad\hfill &\tv\hfill\quad#\qquad\hfill
&\tv\hfill\quad# \qquad\hfill
&\tv#\tabskip=0pt\cr\tablerule   $M_X$ in GeV & $M_I$ in GeV &
  $(k_{1}, k_{2}, k_{3})$
&\cr\tablerule   $10^{18}$&$10^{13}$&$(4,20,18)$
&\tabskip=0pt\cr $10^{17}$&$10^{12}$&$(5,15,12)$
&\tabskip=0pt\cr $10^{16}$&$10^{11}$&$(4,8,6)$
&\tabskip=0pt\cr\tablerule}}

\vskip 1.0truecm
\centerline{\bf Table 4.}
\vskip 0.5truecm

Examples of values of Kac-Moody levels $(k_{3}, k_{2}, k_{1})$ of
$SU(3)_c\times SU(2)_L\times U(1)_Y$ allowing for $SU(4)_c \times
SU(2)_L \times U(1)_R$ at an intermediate scale and charge
quantization.

\vfill\eject
\end